# Computer-readable Image Markers for Automated Registration in Correlative Microscopy – "autoCRIM"

J. Sheriff[1*], I.W. Fletcher[1], P.J. Cumpson[2]

1. School of Engineering, Newcastle University, Newcastle upon Tyne, NE1 7RU, UK

2. Mark Wainwright Analytical Centre, UNSW, Sydney NSW 2052, Australia

* Corresponding Author

E-mail address: j.sheriff2@newcastle.ac.uk (J. Sheriff)

## Abstract

We present a newly developed methodology using computer-readable fiducial markers to allow images from multiple imaging modalities to be registered automatically. This methodology makes it possible to correlate images from many surface imaging techniques to provide an unprecedented level of surface detail on a nanometre scale that no one technique can provide alone.

This methodology provides the capability to navigate to specific areas of interest when transferring samples from machine to machine seamlessly. Then taking data acquired from Scanning Electron Microscope (SEM), Secondary Ion Mass Spectrometry (SIMS), x-ray Photoelectron Spectroscopy (XPS), Atomic Force Microscopy (AFM) and optical inspection tools and combining all the data acquired to then generate a 3D data representative model of a surface.

**Keywords**

1. Correlative Microscopy
2. Surface Analysis
3. Computer Readable Image Marker (CRIM)
4. Automation
5. Image Registration
6. Image Alignment

**Abbreviations**

SIMS – Secondary Ion Mass Spectrometry

XPS – X-ray Photoelectron Spectroscopy

SEM – Scanning Electron Microscope

AFM – Atomic Force Microscopy

CRIM – Computer Readable Image Marker

CLEM – Correlative Light and Electron Microscopy

CM – Correlative Microscopy

EM – Electron Microscope

EDX – Energy Dispersive X-ray Spectroscopy

EXIF – Exchangeable Image File

FIB – Focused Ion Beam

QR – Quick Response

AR – Augmented Reality

PNG – Portable Network Graphics

UNICORN – Universal Image Correlation

1. **Introduction**

Any single imaging modality is unable to provide a full understanding of a surface from the top monolayer down to the bulk, one example is the inner workings within a cell. As the techniques of microscopy proliferate and develop, it is increasingly the case that no single microscopy technique can provide all of the desired information. If it is possible to take advantage of multiple imaging instruments spanning a range of spatial scales and information content, preferably with some of them encompassing preservation of the hydrated nature of the cell. Such a multi-technique approach can capture a significant number of cells to identify features of interest, and perhaps overlay high-resolution snapshots that represent bona fide cellular events.

Traditionally correlative microscopy (CM) is performed by correlating images from at least two different microscopy methods[1], using separate instruments, often in the same laboratory but potentially at separate locations. An example[2] might be the overlaying of a fluorescence microscopy image on one acquired in an electron microscope (EM), but there are many other possible combinations or requirements. A slow step in CM is the laborious process of sample coordinate transformation between different microscopes by finding features common to multiple images, hence the appeal of hybrid systems that combine two separate microscopes in one instrument. The most widely used hybrid system is probably that of an atomic force microscope (AFM) integrated with an inverted light microscope. In this case, the light microscope serves as an excellent tool for identifying cells or other structures using transmitted light and fluorescence microscopy combined with a host of biophysical measurements derived from AFM.

## 2. The problem of registration

Being able to overlay an image from one microscope on an image from another depends on knowing points that each image have in common. Even if the scaling of the images is different, it is generally possible to map an image onto another if one knows some points that occur in both. Sometimes these points can be features introduced into the sample as markers, sometimes one relies on adventitious features of the sample that stand-out well in all imaging techniques.

Since the manual registration of one image compared to another is often uncertain, that approach to overlaying information from different microscopes often results in procedures that can be time-consuming and require a high level of expertise. The problem of co-registration of images can be aided using *fiducial markers*. The process of overlaying one image on another is then reduced to projective transformation of one of the images so that the fiducial markers of both coincide.

## 3. Fiducial Markers

A fiducial marker is an object placed in the field of view of an imaging system which appears in the image produced, for use as a point of reference or a measure. Fiducial markers introduced intentionally as coordinate references in Correlated Light and Electron Microscopy (CLEM) have included scratches, particles and fluorophores. Alignment of an image taken with one microscope and overlaying it with another image taken with another microscope can be a skilled task, even when good fiducial markers exist. The most precise localization is often required for gleaning insights to proteins with unknown function[3].

Another field in which fiducial markers have been important is in robotics, where computer-readable fiducial markers have been used to allow robots to identify and locate objects (including other robots) within a 3D scene using one or more 2D images. Though this may seem a very different field of research to microscopy, one shall see that it can be possible to leverage some of the excellent work already done on this. Fiducial markers make it easier for autonomous robots to track the location and identity of multiple objects in a scene and calculate angles and distances to such objects. Several types of fiducial marker have been developed over the last 15 years or so, largely in the discipline of robotics, with the aim of being easy for a computer to interpret from an image. These include QR codes[4] (familiar from use in posters and consumer products), AprilTags[5], ARTags[6], and circular dot patterns[7]. There has been something of a competition between different computer-readable fiducial marker systems to demonstrate which has the best performance. In this context, performance comprises aspects such as having the lowest false positive rate, lowest false negative rate, and lowest inter-marker confusion rate. These require highly specialized detector and decoder algorithms to find them in images[8], but software of an excellent quality and stability has been developed, often open source.

Of the above-mentioned fiducial marker systems, people are perhaps most familiar with QR-codes. Notice that these are usually used to convey information, such as appearing in an advertisement containing the URL for the advertiser's website. One might point one's smartphone at this QR code to avoid typing in a long URL. The software in the smartphone decodes the small number of bytes of information contained in the URL code that appears in the field of the smartphone camera. The main purpose here is extracting information from the code itself. The position of the QR code within the image, its size or angle of orientation is irrelevant to the purpose of the software. This smartphone software will need to identify the corners of the QR code, and implicitly its size, position and orientation within the image in order to extract the information that the QR code carries. Meanwhile, the software discards the information it has had to extract from the image on the size, position and orientation of the QR code. They are analogous to nuisance parameters in a least-squares fit to some experimental data. In contrast to this kind of scanning QR codes with your smartphone, a fiducial marker has a different purpose. In robotics applications – and is shown in CLEM applications – these geometrical coordinates and angles are very important, and indeed are the reason for using the fiducial codes at all. Therefore, there is a distinction between codes that are primarily designed for carrying information (e.g. QR codes) and those primarily designed for extracting location and geometrical information about the surface on which the code lies (e.g. Apriltags). Superficially these codes look somewhat similar, and each carries some information and could potentially be used for location too. Nevertheless, choices made in the design of these fiducial codes can make it easier for software to identify and locate an Apriltag within an image at distance (say) compared to a QR code. QR codes are optimised for extracting information from an image, Apriltags, ARtags and others are optimised for conveying location from an image (Figure 1).

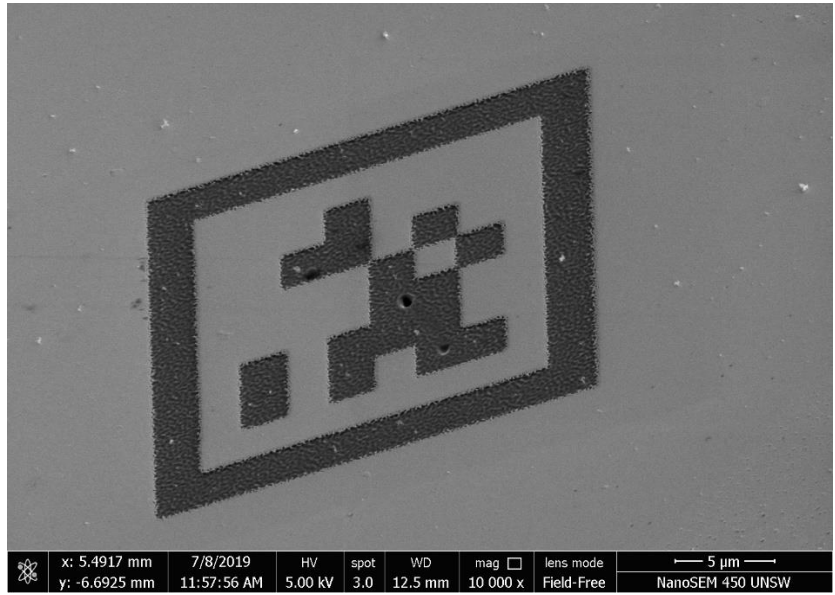

*Figure 1 SEM Images of an Apriltag on Gold coated Silicon wafer*

In microscopy (unlike robotics) the fiducial markers may be designed for human identification only – by manually aligning fiducial marks that appear in more than one image when overlaying them. A scratched cross made with a clean scalpel may be the earliest example of such a human-readable fiducial marker – obvious in both light and electron microscopy images and defining the position of a particular point on a sample. Such a crosshair marking is good for humans, but it can be suggested that there are some good reasons for microscopists to adopt computer readable fiducial markers that have previously been found essential in robotics.

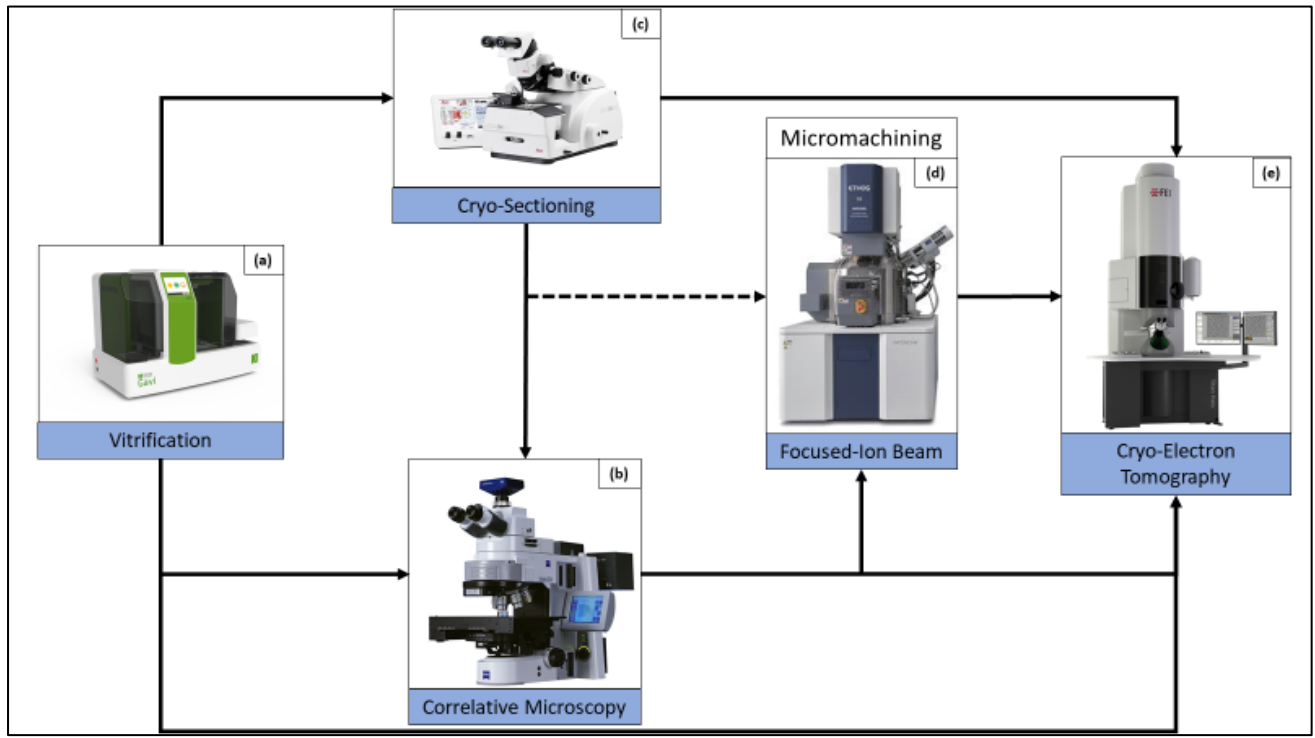

*Figure 2 Route to thin cellular samples in cryo-electron tomography. The schematic diagram depicts the various approaches in sample preparation for cryo-electron tomography. After vitrification of the sample (a), either by plunge freezing or by high-pressure freezing, complex samples are navigated by correlative microscopy (b) and identified areas of interest can be further processed by vitreous sectioning (c) or focused ion beam micromachining (d) for subsequent tomographic analysis in the electron microscope (e).*

## 4. "Proprietary" verses "Open" solutions to the registration of CM images

In recent years some manufacturers have identified the value of CM workflows to their customers and have provided proprietary CM solutions for their instruments (Figure 2). These comprise proprietary means of registering one image with another, typically by one of three methods:

i. Combined instruments (e.g. AFM and inverted microscope) having a common stage coordinate system;

ii. Software solutions: algorithms that find adventitious common features in images from different microscopes;

iii. Combinations of hardware and software that can be used with different microscopes from the same manufacturer. For example, special microscope stages with fiducial markers recognised by the operating software.

Proprietary solutions can be very convenient if one can afford them, because they make it easy to carry-out particular CM workflows if one has all of the equipment from one manufacturer. However, they have two key disadvantages:

(a) They are typically limited to the microscopes offered by the manufacturer concerned. If the manufacturer offers AFM and SEM for example, then even if one is fortunate enough to have bought these instruments from this manufacturer one has no way to apply the proprietary CM method to fluorescence microscopy images of the same sample if the manufacturer does not make fluorescence microscopes;

(b) Commercial priorities change, and the CM solution offered by the company last year may differ to that offered next year.

**5. Requirements for an automatic registration system**

A CM registration system should be:

(a) Applicable to imaging systems at a wide range of length-scale, from images having sub-nanometre resolution to those having millimetre resolution;

(b) Useable with a wide range of existing microscopes and imaging modalities. New microscopes should not be necessary;

(c) As automatic as possible, i.e. requires little or no operator skill in registering one image with respect to another. For example, a microscopist manually lining-up crosshairs present on multiple images is impractical at the throughput of modern microscopes;

(d) To easily be able to find areas needed for analysis by other laboratories so that "high end" techniques can be used by anyone;

(e) Likely to be still around in 10 or 15 years, guaranteeing the continued value of investing in the CM registration system when developing new workflows.

Items (b) and to a large extent (e) provide strong motivation for an ***open*** rather than a ***proprietary*** CM registration system. This means one that relies only on open-source software (and if required, hardware).

### 6. CM Image Registration software running in the background

Our aim is a CM registration system composed of:

i. Computer-readable fiducial markers present within the field of view of some or all images stored;

ii. These images then being stored in a particular location or directory structure;

iii. Open-source software that operates in the background of a PC, going through the contents of specified directories of images, identifying computer readable fiducial markers in each image and attaching the coordinate information to those files. This does not have to be done in real time, indeed could be completed slowly as computer resources are available, e.g. overnight;

**iv.** A user interface, again open source, allowing one to overlay any two or more images selected by the user. This is where projective geometrical transformation is applied.

## 7. Image inversion

Although the work in robotics on computer-readable fiducial markers is very similar to what CM practitioners can use, the two applications are not identical. In robotics, a black-and-white image will never be inverted by a camera, so robots never need to deal with the wide variety of contrast mechanisms (even inversion) that are common in microscopy. Therefore, as well as normalising brightness and contrast prior to passing an image to the Apriltag identification code it is good practice to send that code a normal and an inverted version of the same image (Figures 3,4). One of the two will work, provided some contrast mechanism of some kind exists.

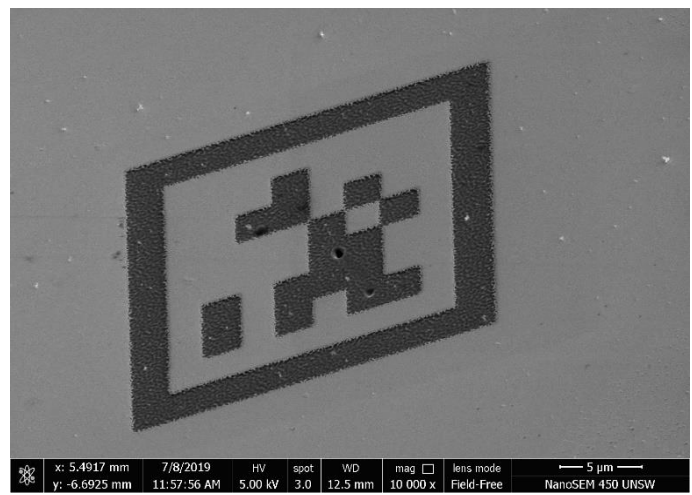

*Figure 3 Not recognised*

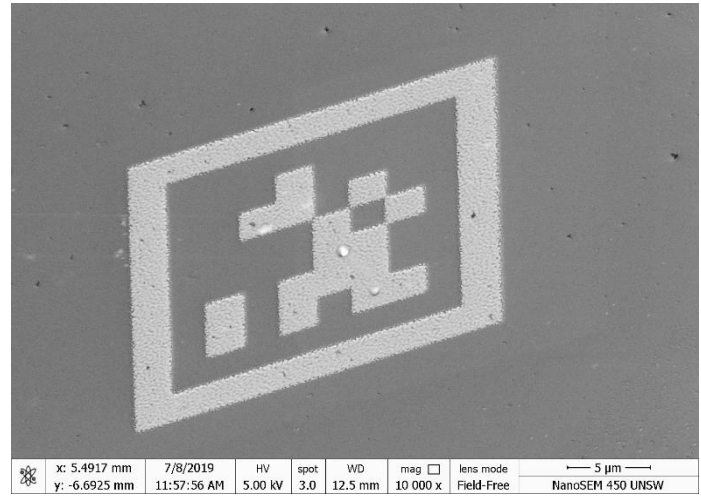

*Figure 4 Inverted image is recognised easily*

It may be sensible to look at other types of contrast mechanism too. For example, edge effects in SEM may be used to advantage and disadvantage.

FIB milling of these computer-readable fiducial markers works well but depends, (of course), on a FIB instrument being available to those performing correlative work. Ten years ago, this would have been uncommon. However, in recent years the need for FIB specimen preparation for TEM means that many more FIB-SEM instruments have been installed and are accessible within a "shared-access" context[9,10,11]. Therefore, one can anticipate that FIB patterning of computer-readable fiducial markers is a good deal easier than previously.

## 8. Co-registration is simplified by having a square fiducial marker

First consider the coordinate system of the FIB instrument that patterns the square tag on the sample surface. For simplicity let us take the size of the square tag to be unity in each dimension. The true size of the square may be 10µm or 50nm for example, but let us take the side of the square tag as twice the unit of distance across the sample surface.

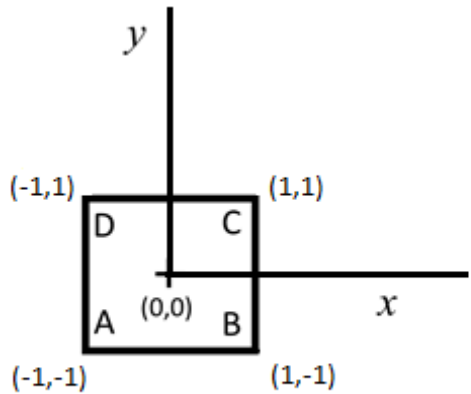

*Figure 5 Representation of an square fiducial marker on the x,y coordinate plane*

For a square fiducial marker, such as an AprilTag or QR code, it is possible to form an image of the position of the centre of the tag, and its four corners (A,B,C and D) as shown in Figure 5. More than one software package has been made available for Apriltag location, and very many for QR code location. These typically provide, at minimum, the location within the pixel array of the four corners of the square tag and its centre. In some cases there is more information available from software that automatically identifies these markers, such as a transformation matrix that would assist in co-registration of images. In our experience, however, these additional matrices are sometimes not available without extra calibration, or tend to be the most unreliable aspects of the output of these software systems because they are not frequently used and therefore not intensively tested. Therefore, proceeding with the aim of using the corners and centre of the fiducial marker alone for image co-registration.

The corners (here labelled A, B, C and D) are distinguishable from each other. The x,y plane of the image is made up of a large number of pixels in a grid, so that for example the corner A of the tag in this image is reported as a pixel having position $x_1'$ and $y_1'$ (Figure 6). Usually these coordinates will be integers, but in some cases the software identifying the tag can give more precise positions as floating-point numbers, for example where the corner can be found to be between pixels.

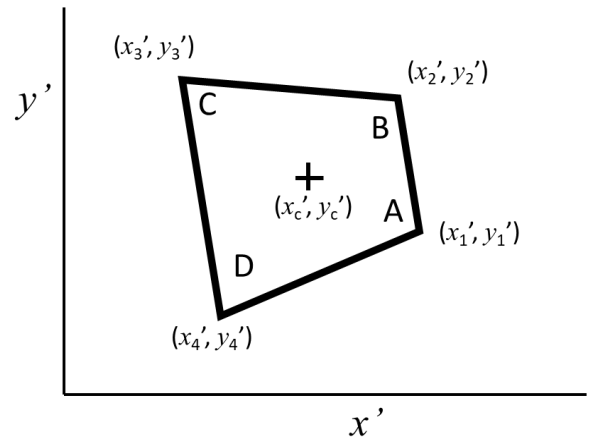

Figure 6 Representation of a square fiducial marker in a different orientation

There are two classes of linear transformations - projective and affine. In most microscopy-relevant transformations an affine description is sufficient, but here the projective model is chosen as it copes better with some oblique views of a surface. The projective transformation can be represented with the following matrix:

$$\begin{pmatrix} a1 & a2 & b1 \\ a3 & a4 & b2 \\ c1 & c2 & 1 \end{pmatrix} \quad \text{Eqn (1)}$$

Where:

$$\begin{pmatrix} a1 & a2 \\ a3 & a4 \end{pmatrix} \quad \text{Eqn (2)}$$

is a rotation matrix. This matrix defines the kind of the transformation that will be performed: scaling, rotation, and so on.

$$\begin{pmatrix} b1 \\ b2 \end{pmatrix} \quad \text{Eqn (3)}$$

is the translation vector. It simply moves the centre of the square, and

$$(c1 \quad c2) \qquad \text{Eqn (4)}$$

is the projection vector. In most applications in microscopy we would expect the elements of this projective vector to be small, though in cases of perspective in macroscopic photography, for example, they can be significant.

If x and y are the coordinates of a point, the transformation can be done by the simple multiplication:

$$\begin{pmatrix} a1 & a2 & b1 \\ a3 & a4 & b2 \\ c1 & c2 & 1 \end{pmatrix} \times \begin{pmatrix} x \\ y \\ 1 \end{pmatrix} = \begin{pmatrix} x' \\ y' \\ 1 \end{pmatrix} \qquad \text{Eqn (5)}$$

Here, x' and y' are the coordinates of the transformed point.

Let's define A to be the transformation matrix;

$$A = \begin{pmatrix} a1 & a2 & b1 \\ a3 & a4 & b2 \\ c1 & c2 & 1 \end{pmatrix} \qquad \text{Eqn (6)}$$

There are five points (the corners and centre of the square tag) that are at known positions in the FIB x,y plane; (x,y)=(-1,-1), (1,-1), (1,1), (-1,1) and (0, 0), this last point being the centre of the tag.

$$A \times \begin{pmatrix} -1 & 1 & 1 & -1 & 0 \\ -1 & -1 & 1 & 1 & 0 \\ 1 & 1 & 1 & 1 & 1 \end{pmatrix} = \begin{pmatrix} x'1 & x'2 & x'3 & x'4 & x'c \\ y'1 & y'2 & y'3 & y'4 & y'c \\ 1 & 1 & 1 & 1 & 1 \end{pmatrix}$$

The values for the five pairs of x' and y' are known from the tag identification in the image. First A has to be found for that image so that we can map the pixels of that image onto the FIB-defined plane (x,y). There are more than enough measurements; indeed this system of equations is overdetermined, and it can be solved in the least-squares sense using the Moore-Penrose pseudoinverse of the (3x5) matrix of (x,y) values in the FIB defined plane. Indeed, since this is always the same, it can be stated that the Moore-Penrose pseudoinverse once-and-for-all as;

$$P = \begin{pmatrix} -5 & -5 & 4 \\ 5 & -5 & 4 \\ 5 & 5 & 4 \\ -5 & 5 & 4 \\ 0 & 0 & 4 \end{pmatrix} / 20$$

So, to find the co-ordinates in the FIB plane of a pixel at (x', y') in any image acquired by a different technique, just multiply those coordinates by the pseudoinverse;

$$\begin{pmatrix} x \\ y \\ 1 \end{pmatrix} = P \times \begin{pmatrix} x' \\ y' \\ 1 \end{pmatrix}$$

## 9. Image imperfections

Clearly, if automatic fiducial markers are to be useful it must be possible for software to find them even in relatively imperfect images. There must be a certain amount of redundancy. This includes imperfect viewing geometry, imaging artefacts and particulate contamination (Figures 7,8).

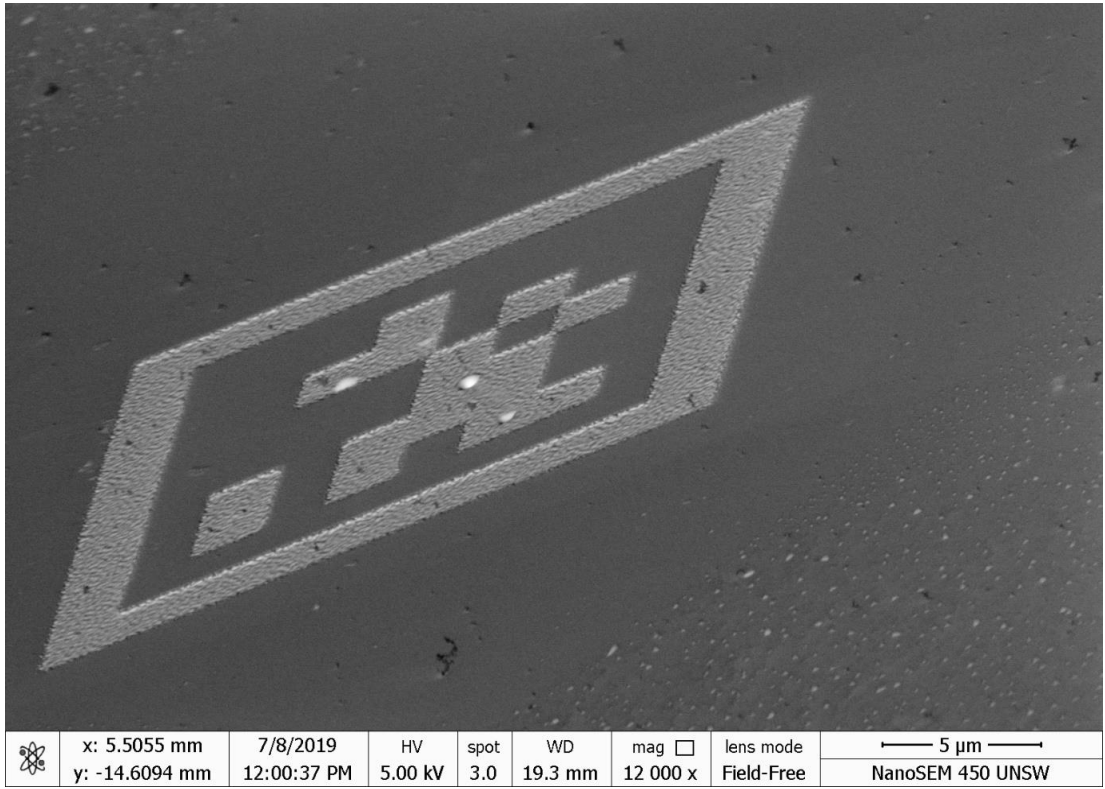

*Figure 7 Imperfect image. Even this oblique image is perfectly reliably recognised, and its orientation quantified*

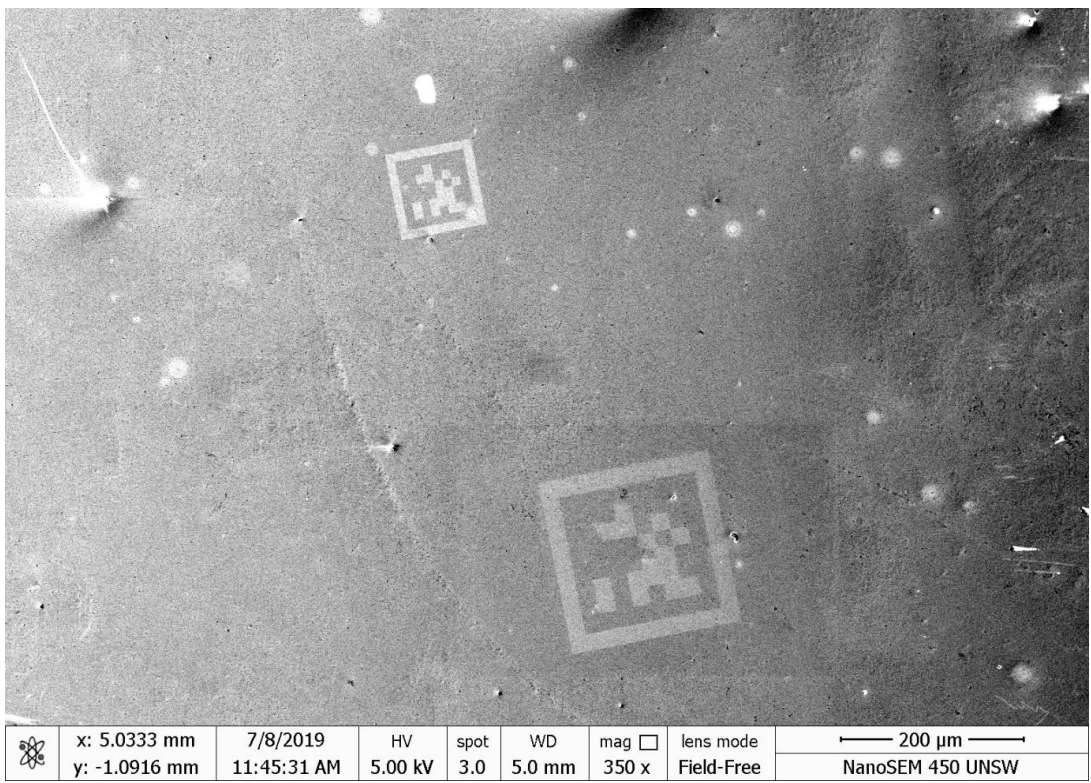

*Figure 8 In this image the lower right Apriltag is recognised easily, but the smaller top-left tag is not - probably due to the algorithm being confused by the light blemish at the lower right corner of this smaller tag.*

## 10. Methodology

### 10.1 Overview of autoComputer-Readable Image Markers (autoCRIM).

As stated previously the task of generating correlative images with different techniques requires specialised sample holders or integrated techniques for example SEM/EDX. While these methodologies have their advantages, they all come with disadvantages too. Specialised sample holders are generally manufacturer specific to the equipment the manufacturer makes so they are limited to where they can be used.

When designing the computer-readable image marker (CRIM) it should be taken into consideration that most operators will only have an optical microscope to align their images with. It is important that the CRIM can be seen optically but also must be chemically different or it would not be visible to certain techniques. When viewing images chemically one will need two different elements, preferably electrically conductive, as this will speed up the calibration of the stage as one only needs to find the CRIM and not worry about charge compensation or other factors. There are many different ways to fabricate CRIMs from micro-milling to photolithography or simply coating a Silicon wafer with Gold then using a Gallium focused ion beam to etch the Apriltag into the Gold layer. This is the method used to generate the images within this paper and is a very quick and easy way of generating the CRIM. It should also be taken into consideration the smoothness of the surface the Apriltag will be applied to as a rough sample can lead to preferential sputtering which can make the tag unreadable.

Now that the CRIM is fabricated all that is left is to apply it to the sample mount which can be a standard 12.5mm SEM stub or another sample holder. If both the

CRIM and sample are kept on the same holder this can be easily transferred to multiple instruments. By making the sample holder part of the sample one can now use the CRIM to calibrate both the stages and techniques. This allows an analyst to easily find areas of interest on a sample either using optical or SE images, then transfer the sample to another instrument or another analyst to carry out another surface analysis technique. This can be achieved by a 2-stage process, first is stage calibration and optimisation and the second image calibration and organisation which is outlined in Figure 9.

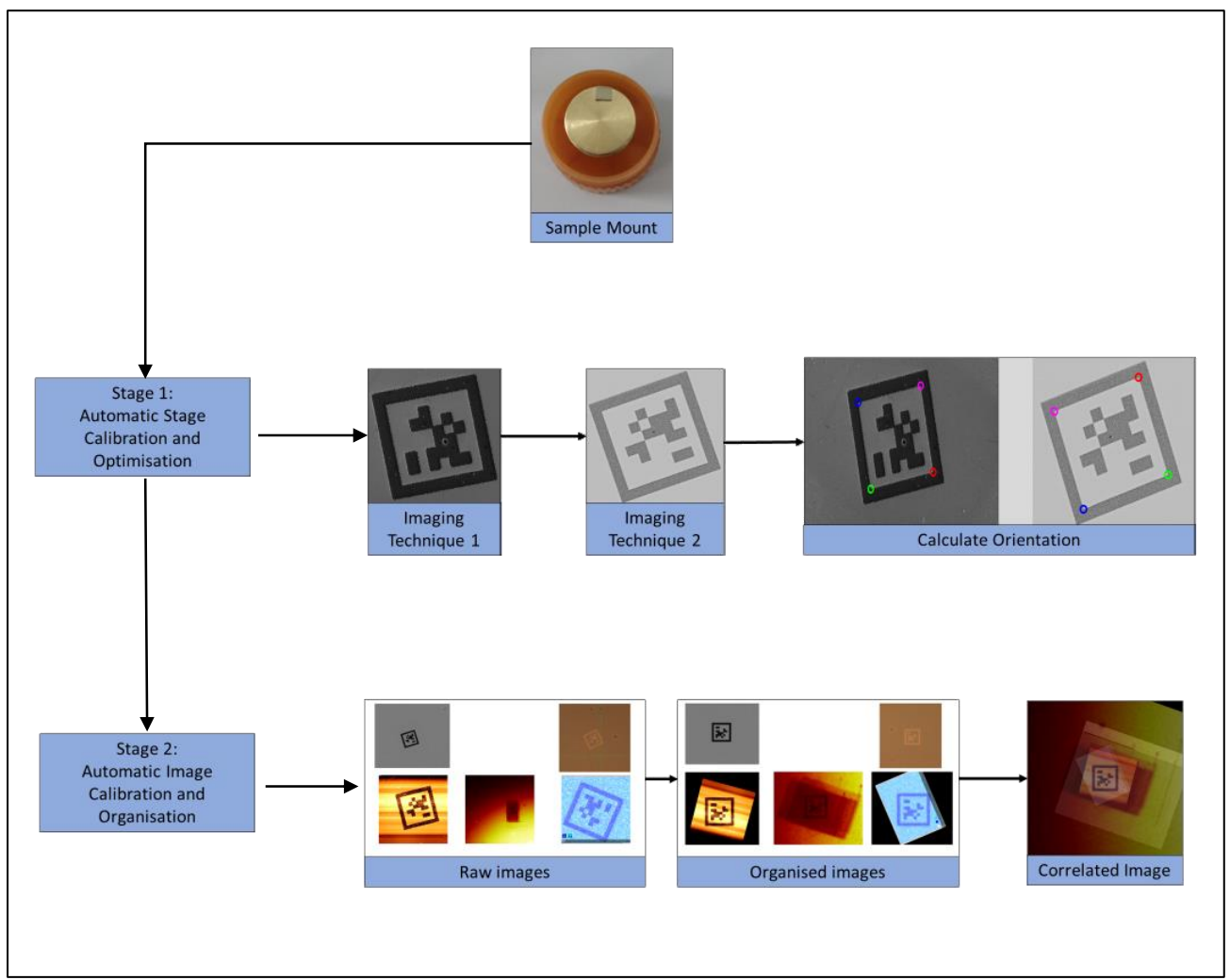

*Figure 9 A workflow of the autoCRIM method showing Stage 1 the locating and orientation of the same CRIM after transfer to a second imaging technique. Stage 2 shows the identification of the CRIM and adjusting the CRIM to be in the centre of the image to finally build the 3D model of the surface.*

**10.2 Focused Ion beam milling**

To achieve a tag that has two chemically distinct areas there are a few considerations that should be known when designing a tag for milling, including the sample analysis depth of the techniques to be used as well as the spatial size of the tag features. The following will outline how the tags were created and certain factors that should be taken into consideration with the design of the tags.

The thickness of the second layer will be dictated by the sample analysis depth of the surface technique intended to be used for example, with SIMS which has a typical analysis depth of 1-5nm the gold layer must be greater than 5nm. 25nm was the chosen thickness to make sure that the gold layer was uniform after the sputtering coating process. To achieve a flat surface a silicon wafer mount was used which was then cut to a smaller size. A Polaron SC7640 was used to sputter 25nm of gold onto the silicon wafer.

The spatial feature size is dictated by the bitmap images of the generated tags which can be found at https://github.com/AprilRobotics/apriltag-imgs. Each tag is a 10x10 pixel image. Now by applying a size to the tag, for example 50µm, each pixel will be a 5µm² area which scales with the overall size of the tag. To decide on a good spatial feature size, we must look to the instrument spot sizes that are to be used. XPS imaging has an optimum spot size of 3µm so if the features are smaller than the spot size the tag will not be seen. Also, if the tag is 500µm and the field of view of the instrument the is 250µm the tag will not be fully imaged so the techniques cannot be correlated with that tag.

A delicate balance must be found between overall size and total etching time. While tags of 25, 50, 100, 250 and 500µm were created the smallest and largest tags were not fit for purpose. While size can vary depending on what technique is required a recommended tag size would be 100µm as it can be seen easily optically and also fits fully into most instruments field of view.

All of the tags created for this paper come from the tag36h11 family. These images were then converted from '.PNG' format to bitmaps so they could be imported into the FIB software. There are now a wide variety of FIB systems with their own patterning software. The etching of these tags was carried out by a Gallium focused ion beam attached to a Zeiss Orion Nanofab. This works in a similar way to typical FIB-SEMs with the FIB attached at a 36° angle from the primary column. The NanoPatterning and Visualization Engine (Zeiss NPVE version 4.7) software was used to etch the tag. A current of 100pA was recorded and the pattern etched for 20 minutes with a spot size of 600nm.

While the GaFIB can achieve spatial resolution higher than what has been used in this work it can be sacrificed to allow for larger currents to decrease the overall etching time. Care must be taken to make sure the spatial resolution is never close to or lower than the spatial features of the tag. It should also be noted that with ion etching there is an interaction volume in the sputtered material meaning that the etched areas will be slightly larger than the un-etched areas. This problem is only an issue for smaller tags or large spot sizes. Currents can also be increased to decrease the overall etching time but a target of >0.07 nC/µm² ion dose should be achieved for the removal of 25nm of Gold.

**10.3 Stage 1 – Automatic Stage Calibration and Optimisation**

The first part of our autoCRIM method involves calibration of the stage to a known set position. Unlike previous correlative methods where one would use a calibration grid to find the area of interest needed for analysis, this method uses an adaptable CRIM that can be attached wherever the analyst desires to allow for more freedom and less restriction that a calibration grid causes. The CRIM is then used to create an origin point on the sample which will allow an analyst to then calibrate each technique to be homed at the same position on the sample holder. As shown in Figure 10 A-B the CRIM can be found very easily with an optical microscope. Once the CRIM and sample have been mounted to the desired holder the first task is to find the CRIM then the areas of interest, this is best done with either an optical or electron microscope depending on the scale of the task. Once the analyst has found the CRIM they can focus and optimise the beam while also setting the correct analysis height for their instrument using the CRIM as a reference due to its distinctive and conductive nature. Once the analyst has taken the image of the CRIM using preferably either an optical or electron microscope technique, they can now look for their areas of interest, which can be as many or as few as required. It is also good practice for insulating samples to add an area that can be used to tune up the charge compensation before targeting the area of interest in order to minimise sample damage. After all of the areas of interest have been imaged the analyst can then add these to the Universal Image Correlation (UNICORN) program that was designed alongside the CRIM's to be able to detect them. The UNICORN will then take the stage coordinates and working distances from the EXIF data (Exchangeable Image File) from the images or can be inputted manually dependant on the analyst preference. From here UNICORN will take the CRIM as the origin point and then calculate the coordinates of all the other areas of interest relative to the CRIM reference position. Now the sample with CRIM needs to be transferred to the next imaging technique and once there the CRIM must be found again. Once the tag has been found the analyst can carry out the imaging analysis on the CRIM. Due to the design of the CRIM it only takes a couple of minutes to locate it and then all that needs to be done is to optimise the conditions for the imaging of the CRIM. Figure 10 C-H shows various examples of imaging techniques of the CRIM. Once the image has been collected it is then binarized using a defined thresholding. The two images are then passed to the detection software: the original binary and the inverted version to detect the tag. When comparing the 2 images of the same CRIM the orientation can then be calculated between the two CRIMs. Once this is calculated, one can apply it to locate where the areas of interest are located relative to the CRIM. Then a list of coordinates can be generated that can be inputted to the next analytical instrument to automatically collect the data or manually if the analyst would prefer. This step can then be repeated depending on how many different surface analysis techniques are needed to find the answer.

### 10.4 Stage 2 – Automatic Image Calibration and Organisation

The second part is to stitch all the images taken from an instrument to form an image set keeping the relative distances from the CRIM and producing one image. UNICORN will search through all the image sets and find the CRIM. Once the CRIM has been found a transform will be applied to the CRIM to place it in a set orientation and centre the CRIM. When this process is finished the images can be then overlayed, Figure 11 was generated using a plugin called TrakEM 2[12] within the image processing software called ImageJ. Now that all the images are overlayed a 'google map' has been created of the sample but here each pixel can now contain not just one set of data. Pixels can now contain mass spectrum data as well as secondary electron intensity, and much more[13]. This will lead to a better understanding of the data as the physical surface can now have chemical data applied to it as well as topography information which will lead to a better understanding of a surface on a nanoscale.

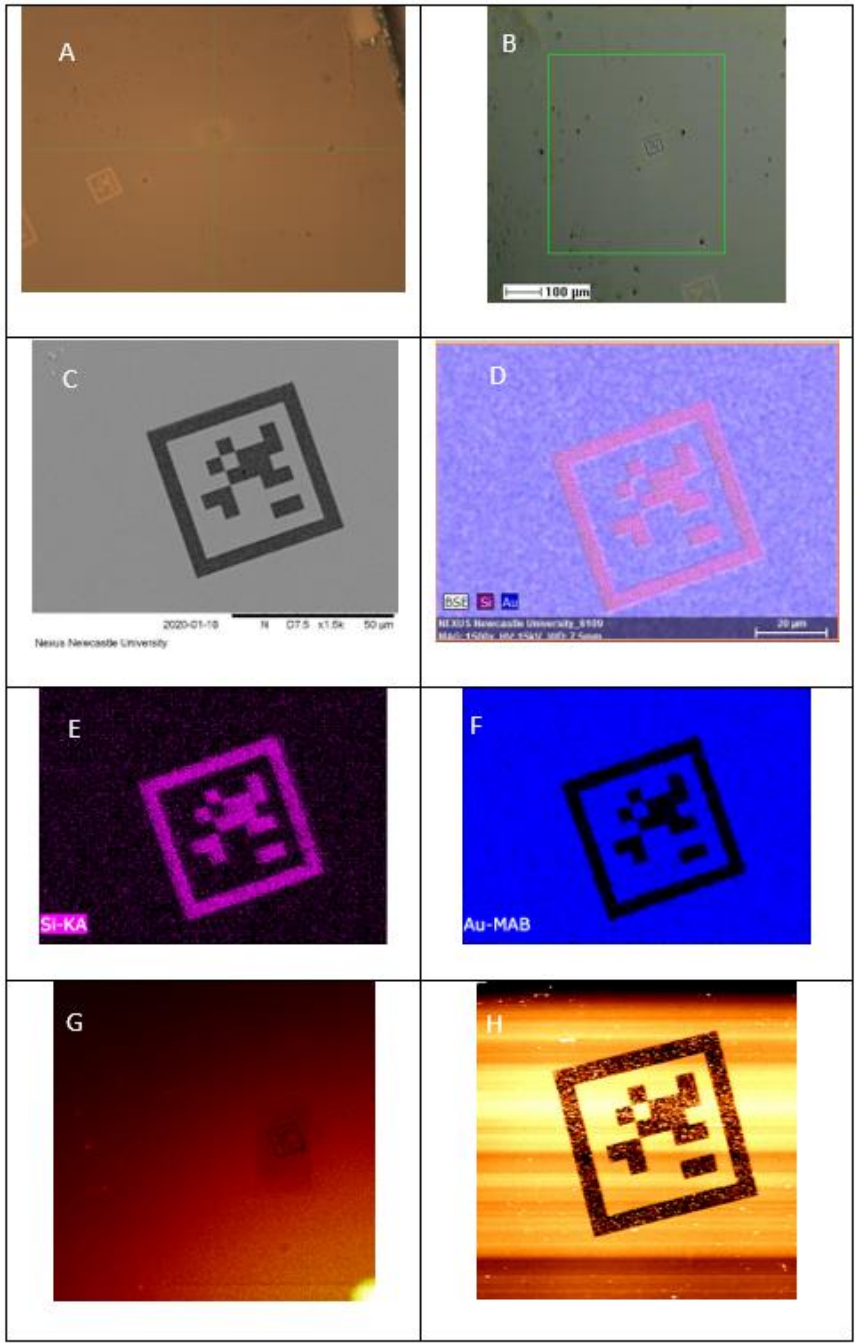

Figure 10 (A) optical image of CRIM using Mitutoyo quick scope. (B) optical image taken with ION-TOF IV camera. (C) Secondary electron image taken with a Hitachi TM3030 tabletop scanning electron microscope at a magnification of 1.5k. (D, E, F) Energy dispersive X-ray spectroscopy using a Bruker XFlash MIN SVE, image D showing the other all signal counts with E showing only Silicon and F only showing Gold counts. (G) Total secondary ion image using a ION-TOF IV. (H) Atomic force microscopy image taken using a Park XE-120 Advanced Scanning Probe Microscope.

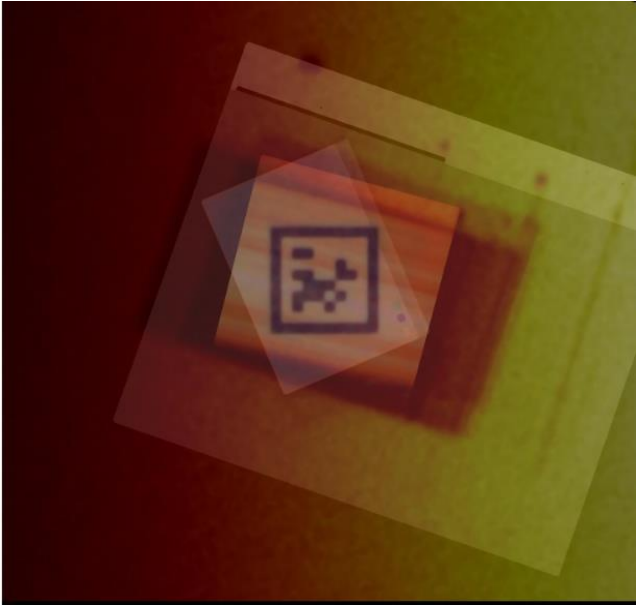

*Figure 11 Overlay of SEM, EDX, AFM and SIMS images of the 50µm Apriltag using TrakEM2*

## 11. Conclusions

We have demonstrated computer-readable fiducial markers for automatic registration of images from multiple microscope modalities. These are facilitated by the wide availability of focussed ion beam (FIB) guns on modern scanning electron microscopes.

The same computer-readable image marker (CRIM) can be easily found and then imaged with a variety of independent surface analytical techniques. These images can be used to overcome the limitation of each technique and give a better understanding of the surface composition and physical characteristics of a surface at the nanoscale.

The nature of the CRIM also removes the need to add surface marker e.g. a grid, or chemical marker to a sample that can alter or damage surface properties that wish to be analysed. These CRIMs can also be transferred between different laboratories as has been demonstrated by the images generated in this paper coming from both UK and Australia.

Further work will be needed to assess the registration accuracy of the CRIM for detection as well as the location of areas of interest. Once this has been carried out the autoCRIM method aims to bring a new standardised method to correlative imaging. This has the ability to quickly create an X, Y coordinate system that can be applied in any surface analysis technique to quickly and accurately find areas of interest when transferring a sample between two or many different techniques.

**Acknowledgements**

We would like to thank all the staff and students at Nexus. Thanks to Mr Mike Foster for providing a high level of support this project. Thanks to Dr Jose Portoles and James Hood for their understanding and advice in programming. I would like to thank my supervisors Professors Grant Burgess and William Willats at Newcastle university and P&G especially Dr Eric Robles who gave me this opportunity.

**Funding Sources**

This work was supported by the "EPSRC and P&G for a CASE PhD studentship" and "EPSRC for equipment funding under the "great eight" scheme and the EPSRC Mid-Range Facility in X-ray photoelectron spectroscopy 2011-2016.